\UseRawInputEncoding
\documentclass{optica-article}

\journal{opticajournal} 

\articletype{Research Article}

\usepackage{lineno}

\begin{document}

\title{Cladding effect on the mode index engineered tuned cavity}

\author{Mohit Khurana,\authormark{1,2,*}}

\address{\authormark{1} Department of Physics and Astronomy, Texas A\&M University, College Station, TX 77843, USA\\
\authormark{2}Institute of Quantum Science and Engineering, Texas A\&M University, College Station, TX  77843, USA}

\email{\authormark{*}mohitkhurana@tamu.edu} 


\begin{abstract*}Photonic integrated circuits require a cladding material on top to prevent any outside interaction with the photonic circuit elements and electromagnetic modes that could cause damage. Mohit et al. \cite{mohit_2024a} proposed selective tuning of the resonance of a cavity by using the mode-index engineering method. However, their work did not consider adding a cladding (upper cladding) material on top of the tuned cavity. In this study, I aim to build upon their work by investigating the impact of depositing cladding material on the frequency distribution of tuned cavities through analytical studies and numerical experiments. I identify crucial calculation parameters and discuss the optimum conditions for high-resolution tuning and large tuning range.
\end{abstract*}
\\
\\

\noindent
For the total length of cavity L and the effective mode index $n$, the resonance frequency is given by $f = \frac{mc}{Ln}$, where m is the mode number and c is the speed of light in vacuum. When the tuning process is performed by the mode-index engineering method \cite{mohit_2024a}, the resonance frequency is tuned to $f'$. The tuned resonance frequency is given by $f' = \frac{mc}{(L-L')n + L'(n+\delta n)}$, where $ \delta n = n' - n$, $L'$ is the partial length of cavity whose effective mode index $n$ is changed to $n'$ and $L-L'$ is the length of the cavity with unchanged effective mode index $n$ and L is the total length of the cavity.  The relative change in resonance frequency is evaluated by, $\frac{\delta f}{f} = \frac{f'-f}{f} =\frac{1}{1 + \frac{L'\delta n}{Ln}} - 1$.

\begin{figure}[ht!]
\centering\includegraphics[width=15.5cm]{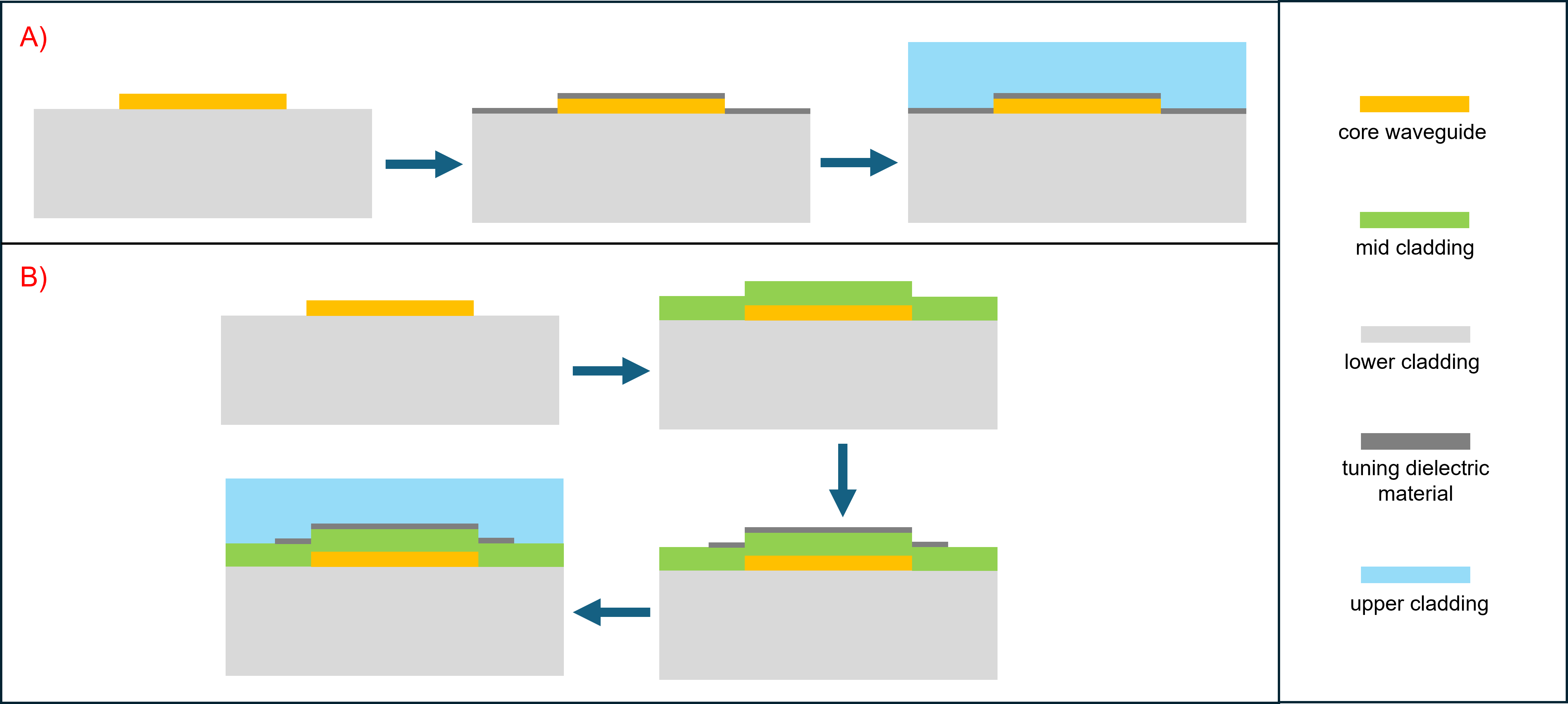}
\caption{The schematic diagram of the layer structure cross-sections of two configurations. First configuration (A): cavity core waveguide material $Si_{3}N_{4}$ (yellow color) on top of $SiO_{2}$ cladding (grey color). A dielectric material layer (dark grey color) is deposited on the partial length of the cavity to tune the cavity's resonance selectively. Finally, an upper cladding is added (blue color). Second configuration (B): cavity core waveguide material $Si_{3}N_{4}$ (yellow color) on top of $SiO_{2}$ cladding (grey color) with a mid-cladding (green color) on top of $Si_{3}N_{4}$. A dielectric material layer (dark grey color) is deposited on a partial length of the cavity to tune the cavity's resonance selectively. Finally, an upper cladding is added (blue color).}
\label{Cladding_on_tuned_cavities}
\end{figure}
\noindent
The addition of cladding on top of the tuned cavity changes the effective mode index and resonance shift from $f'$ to $f_{cladding}$. So, the resonance frequency shifts in two steps, 
$$f \rightarrow f' \rightarrow f_{cladding} $$
where $f'$ is tuned cavity resonance.
$$f = \frac{mc}{Ln} \rightarrow f' = \frac{mc}{(L-L')n + L'(n+\delta n)} \rightarrow f_{cladding} $$
For any tuned cavity, $(L-L')n + L'(n+\delta n) = \text{constant} = K$. After adding a cladding on top i.e. upper cladding,
$$\text{effective mode index of L-L' segment changes}:\,\,\, n \rightarrow n'$$
$$\text{effective mode index of L' segment changes}:\,\,\,n + \delta n \rightarrow n_{p}$$
For small values of $(n_{p}-\delta n -n)$, and $(n'-n)$, mode number $m$ remains the same. In Fig. \ref{Cladding_on_tuned_cavities}, two configurations are shown, A and B; in the case of A, $\delta n$ is large and limits high tuning resolution, but in B, $\delta n$, $\delta \alpha$ and $\delta \beta$ are small numbers, high-Q resonators are fine-tuned i.e. configuration B is more favorable for tuning procedure because of the small $\delta n$, $\delta \alpha$, and $\delta \beta$. Therefore, I limit the numerical experiment to configuration B only.  Now I evaluate, $\frac{\Delta f'_{cladd.}}{f'} = \frac{f_{cladding} - f'}{f'}$,
$$\frac{\Delta f'_{cladd.}}{f'} = \frac{(L-L')n + L'(n+\delta n)}{(L-L')n' + L'n_{p}} - 1 = \frac{K}{(L-L')n' + L'n_{p}} - 1$$
I introduce two new parameters $\delta \alpha$ and $\delta \beta$,
$$n_{p} = n + \delta n + \delta \alpha$$
$$n' =  n + \delta \beta$$

\noindent
where $\delta \alpha$ is the change in the effective mode index of the L' segment, and $\delta \beta$ is the change in the effective mode index of the L-L' segment after adding a cladding on the tuned cavity.
$$\frac{\Delta f'_{cladd.}}{f'} = \frac{(L-L')n + L'(n+\delta n)}{(L-L')(n+\delta \beta) + L'(n+\delta n +\delta \alpha)} - 1= \frac{K}{(L-L')(n+\delta \beta) + L'(n+\delta n +\delta \alpha)} - 1$$
$$\frac{\Delta f'_{cladd.}}{f'} = \frac{K}{K + (L-L')(\delta \beta) + L'(\delta \alpha)} - 1$$

\begin{figure}[ht!]
\centering\includegraphics[width=15.5cm]{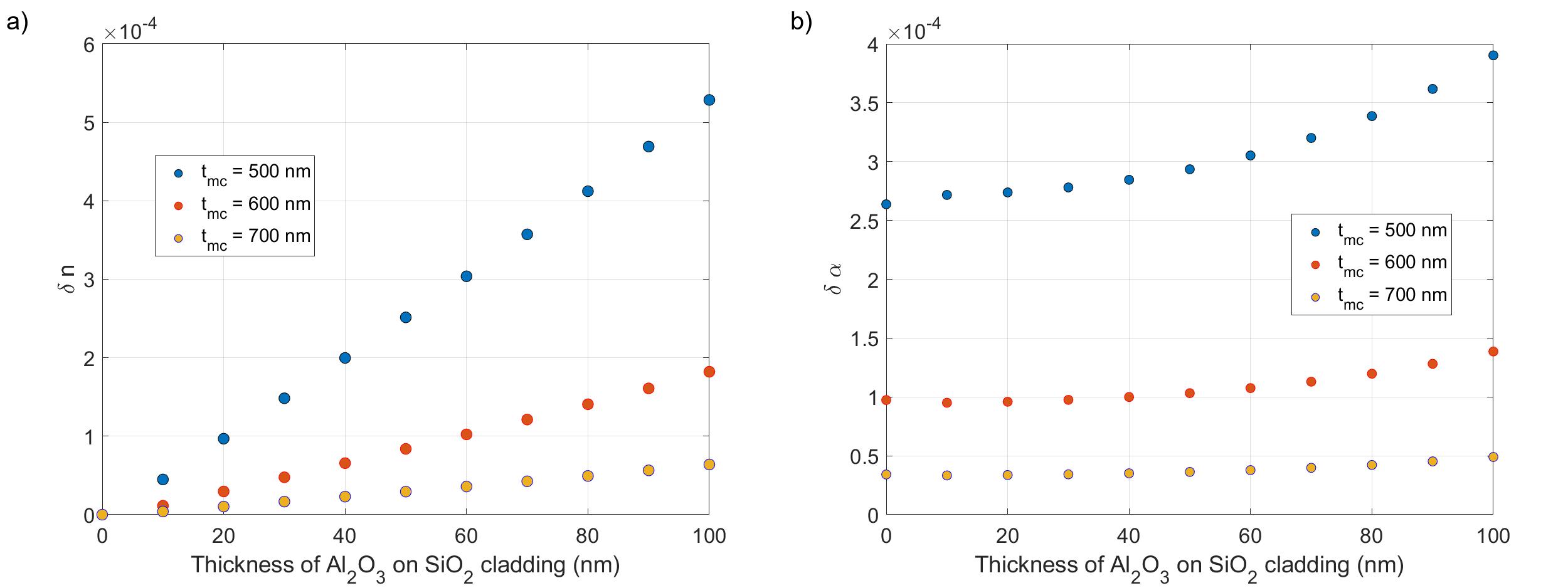}
\caption{a) Plot of $\delta n$ vs. thickness of $Al_{2}O_{3}$ layer on $SiO_{2}$ cladding without an upper cladding. $t_{mc}$ is the mid $SiO_{2}$ cladding thickness. b) Plot of $\delta \alpha$ vs. thickness of $Al_{2}O_{3}$ when an upper cladding $SiO_{2}$ is added on top. $\delta \alpha$ is evaluated by estimating the effective mode index (with upper cladding) and subtracting the effective mode index (without upper cladding). Note that, at zero thickness of $Al_{2}O_{3}$ i.e. $\delta n =0$, $\delta \beta$ = $\delta \alpha$. For both (a) and (b), three different thicknesses of mid $SiO_{2}$ cladding ($t_{mc}$) are considered, i.e., 500 nm, 600 nm, and 700 nm. For the larger value of $t_{mc}$, smaller values of $\delta n$ and $\delta \alpha$ are found which is obvious due to less mode overlap with $Al_{2}O_{3}$ and upper cladding as $t_{mc}$ increases. But the critical parameter is $\Delta M$, which is dependent on $\delta \alpha -\delta \beta$.}
\label{array_a}
\end{figure}

\begin{figure}[ht!]
\centering\includegraphics[width=15.5cm]{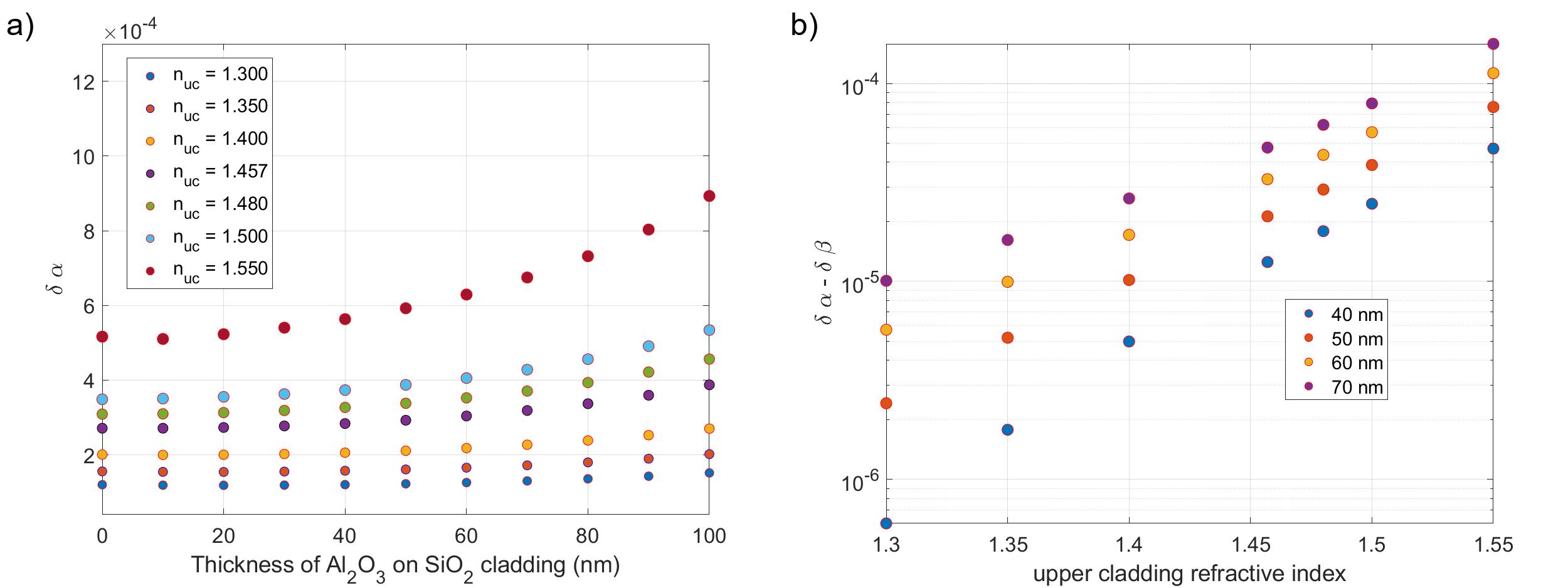}
\caption{Upper cladding refractive index ($n_{uc}$) dependent $\delta \alpha$ and $\delta \alpha - \delta \beta$ for $t_{mc}$ = 500 nm, a) Plot of $\delta \alpha$ vs. thickness of $Al_{2}O_{3}$ on 500 nm $SiO_{2}$ mid cladding and b) Plot of $\delta \alpha - \delta \beta$ vs. $n_{uc}$ for 40 nm, 50 nm, 60 nm, and 70 nm thick $Al_{2}O_{3}$ on 500 nm $SiO_{2}$ mid cladding; $\delta \beta$ is calculated at zero thickness of $Al_{2}O_{3}$ from (a).}
\label{array_b}
\end{figure}

\noindent
Ideally, I want $\frac{\Delta f'_{cladd.}}{f'}$ to be independent of L' so that adding an upper cladding doesn't introduce a worsening of frequency distribution, though the resonance will shift. Only two factors, $\delta \alpha$ and $\delta \beta $, contribute to the shift of resonance and worsening of frequency distribution due to the addition of cladding on top of tuned resonators ($K$ is constant for all tuned cavities).
The critical quantity is, $\Delta M$ which is defined as, $$\Delta M = \left(\frac{\Delta f'_{cladd.}}{f'}\right)_{L'=L}  - \left(\frac{\Delta f'_{cladd.}}{f'}\right)_{L'=0}$$ This evaluates the maximum change in frequency distribution due to addition of an upper cladding.

$$\Delta M =  \frac{1}{(1 +\frac{\delta \alpha}{n +\delta n})} - \frac{1}{(1+ \frac{\delta \beta}{n})}$$



$$\Delta M \sim \frac{ ( \delta \beta -\delta \alpha)}{n + (\delta \beta + \delta n +\delta \alpha)} \sim \frac{ ( \delta \beta -\delta \alpha)}{n}$$
Only the magnitude of $\Delta M$ is relevant. Ideally, I want this quantity to be below 1/Q or tuning resolution. These numbers depend on the cavity's material configuration and should be evaluated using numerical methods such as finite-difference time-domain (FDTD) analysis. For an example of a numerical experiment, I consider a 100 nm thick 4000 nm wide silicon nitride ($Si_{3}N_{4}$) core waveguide on silicon dioxide ($SiO_{2}$) cladding, which I call lower cladding. Another $SiO_{2}$ cladding is on top of $Si_{3}N_{4}$ material, which I call mid cladding, and $Al_{2}O_{3}$ layer is used for tuning step, and the top cladding is called as upper cladding. The refractive index of $Si_{3}N_{4}$, $SiO_{2}$ and $Al_{2}O_{3}$ are assumed as 2.000, 1.457, and 1.76 respectively at 772 nm. The simulation results are estimated at the operating wavelength, 772 nm, for fundamental quasi-transverse electric (TE$_{0}$) mode. Fig. \ref{Cladding_on_tuned_cavities} shows two configurations, A and B. When the thickness of mid-cladding ($t_{mc}$) decreases, the mode overlaps with tuning dielectric and upper cladding material increases, and $\delta n$, $\delta \alpha$, and $\delta \beta$ increase; configuration B becomes A when the thickness of mid-cladding is zero. Fig. \ref{array_a}(a) and (b) shows the dependence of $\delta n$, $\delta \alpha$ on $t_{mc}$ and thickness of tuning dielectric material $Al_{2}O_{3}$, respectively. Note that values of $\delta \alpha$ (see Fig. \ref{array_a}(b)) are similar in order of $\delta n$ when an $SiO_{2}$ upper cladding is added. But when the $t_{mc}$ is large the variation of $\delta \alpha $ is small with thickness of $Al_{2}O_{3}$ or $\delta n$. This is an important observation; a relatively larger value of the mid-cladding and tuning dielectric material thicknesses smaller values are found for $\delta \alpha$ and $\delta \beta$ (see Fig. \ref{array_a}(b), \ref{array_b}(a) and (b)). Fig. \ref{array_b}(a) and (b) shows the $\delta \alpha $ and $\delta \alpha -\delta \beta$ dependence on the thickness of tuning dielectric material $Al_{2}O_{3}$ and upper cladding refractive index ($n_{uc}$) for $t_{mc}$ = 500 nm. Note that as the upper cladding refractive index decreases, $\delta \alpha - \delta \beta$ decreases significantly, favoring the mode index engineering method for tuning applications.
In the discussed example, $\Delta M$ is $\sim 4\times 10^{-7} - 10^{6}$ for $\delta n \sim 2 \times 10^{-4}$ and $n_{uc}$ = 1.3 - 1.35. Therefore, the method discussed by Mohit et al. \cite{mohit_2024a} attains high-resolution tuning and a wide tuning range for wafer-scale manufactured microring cavities with loaded-Q factor in the order of 1 million. In principle, even higher Q resonators can be tuned with appropriate optimizations of materials and use of low-index upper cladding materials.

\noindent
Now, I work out the $\Delta M$ quantity for adiabatic design (see Ref. \cite{mohit_2024a}). For adiabatic tapered regions (a and b; a = ad.1 at one end and b = ad.2 at the second end), 
$$f'=\frac{mc}{(L-L'-2l)n + L'n' + \int_{ad.1} n_{a}(x) dx + \int_{ad.2} n_{b}(x) dx}$$

\noindent
where $L-L'-2l$ region has effective mode index $n$, $L'$ region has effective mode index $n'$ and two regions (a and b) of length $l$ have effective mode index as a function of x, i.e., $n_{a}(x)$ and $n_{b}(x)$. Assuming the tapering region is linear of each adiabatic region of two ends (ad.1 and ad.2),
$$\int_{ad.1} n_{a} (x) dx = \int_{ad.2} n_{b} (x) dx= (n+n')l/2$$
\noindent
and after the addition of cladding, $n_{a}(x)$ and $n_{b}(x)$ becomes $n'_{a}(x)$ and $n'_{b}(x)$ respectively, therefore,
$$\int_{ad.1} n'_{a} (x) dx = \int_{ad.2} n'_{b} (x) dx= (n'+n_{p})l/2$$

\noindent
Therefore,
$$\frac{\Delta f'_{cladd.}}{f'} = \frac{(L-L'-2l)n + L'(n+\delta n)+(2n+\delta n)l}{(L-L'-2l)n' + L'n_{p}+(n'+n_{p})l} - 1$$
The quantity $\Delta M$ is estimated as,
$$\Delta M = \left(\frac{\Delta f'_{cladd.}}{f'}\right)_{L'+2l = L}  - \left(\frac{\Delta f'_{cladd.}}{f'}\right)_{L'=0, l=0}$$

$$\Delta M \sim \frac{\delta \beta- \delta \alpha}{n}$$

\noindent
which is similar to the $\Delta M$ for non-adiabatic design. This is because only the difference of $\delta \beta$ and $\delta \alpha$ contribute significantly to $\Delta M$, and $l$ is significantly smaller compared to $L$.


\section{Conclusion}
Cladding on top of photonic-integrated devices reduces signal loss and crosstalk. It also offers protection against environmental damage, ensuring device reliability and durability. Therefore, it is essential to investigate the effect of thick cladding on top of tuned microring cavity. In this work, I have extended the work reported by Mohit et al. \cite{mohit_2024a} by investigating the impact of upper cladding on tuned cavity. I introduced new parameters and quantities to estimate the worsening of frequency distribution due to cladding deposition on top of tuned cavities. It is found that the mode index-engineering method for passive tuning applications allows a scalable single-step tuning process for high-Q microring resonators. In the discussed numerical experiment, resonators with loaded Q-factor up to $10^{6}$ can be tuned with resolution below 1/Q and tuning range up to $10^{3}/Q$ without any significant impact on the frequency distribution after the addition of low index upper cladding. Low index materials for upper cladding are favorable to minimize the $\Delta M$; this is evident due to the high index contrast between the mid-cladding and upper-cladding material. Therefore, it is essential to review low-index materials that are optically transparent, low-cost, and compatible with complementary metal-oxide-semiconductor (CMOS) technology. The high-quality growth of such low-index materials on top of materials used in the mid-cladding and tuning step is a crucial area for future research. I envision that the work presented here will be expanded in further research, including different materials such as silicon, lithium niobate, silicon carbide, and experimental demonstrations.






\section{Funding}
I want to thank the Robert A. Welch Foundation (grants A-1261 and A-1547), the DARPA PhENOM program, the Air Force Office of Scientific Research (Award No. FA9550-20-10366), and the National Science Foundation (Grant No. PHY-2013771). This material is also based upon work supported by the U.S. Department of Energy, Office of Science, Office of Biological and Environmental Research under Award Number DE-SC-0023103, DE-AC36-08GO28308. 

\section{Acknowledgment}
I want to thank Prof. Marlan O. Scully, who provided continued support and funds to conduct this study.

\section{Disclosures}
The author declares that he has no known competing financial interests or personal relationships that could have appeared to influence the work reported in this paper.

\section{Data Availability}
Data underlying the results presented in this paper are not publicly available at this time but can be obtained from the author upon reasonable request.



\bibliography{sample}

\end{document}